\documentclass[a4paper]{jpconf}
\usepackage{graphicx}
\begin{document}
\title{Hunting keV sterile neutrinos with KATRIN: building the first TRISTAN module}
\author{T~Houdy$^{1,2,*}$, A~Alborini$^{3}$, K~Altenm\"uller$^{2}$, M~Biassoni$^{4,5}$, L~Bombelli$^{3}$, T~Brunst$^{1,2}$, M~Carminati$^{6}$, M~Descher$^{7}$, D~Fink$^{1}$, C~Fiorini$^{6}$, M~Gugiatti$^{6}$, A~Huber$^{7}$, P~King$^{6}$, M~Korzeczek$^{7}$, M~Lebert$^{2}$, P~Lechner$^{8}$, S~Mertens$^{1,2}$, M~Pavan$^{4,5}$, S~Pozzi$^{4,5}$, D~C~Radford$^{7}$, A~Sedlak$^{1}$, D~Siegmann$^{1,2}$, K~Urban$^{2}$, J~Wolf$^{7}$}
\address{$^1$ Max-Planck-Institut f\"ur Physik, F\"ohringer Ring 6, D-80805 M\"unchen, Germany}
\address{$^2$ Physik-Department, Technische Universit\"at M\"unchen, D-85747 Garching, Germany}
\address{$^3$ XGLab SRL, Bruker Nano Analytics, Via Conte Rosso 23, Milano, Italy}
\address{$^4$ Universit\`a  di  Milano - Bicocca, Dipartimento  di  Fisica,   I-20126  Milano, Italy}
\address{$^5$ INFN  Sezione  di  Milano  -  Bicocca,  I-20126  Milano  -  Italy}
\address{$^6$ Politecnico di Milano, Dipartimiento di Elettronica, Informazione e Bioingegneria, Piazza Leonardo da Vinci, 32, I-20133 Milano, Italy}
\address{$^7$ Karlsruhe Institute of Technology, Hermann-von-Helmholtz-Platz 1, D-76344 Eggenstein-Leopoldshafen, Germany}
\address{$^8$ Halbleiterlabor der Max Planck Gesellschaft, Otto-Hahn-Ring 6, D-81739 München, Germany}
\address{$^9$ Oak Ridge National Laboratory, 1 Bethel Valley Road, Oak Ridge, 37831 TN, USA}

\ead{*thoudy@mpp.mpg.de}
\bibliographystyle{iopart-num}

\begin{abstract}
The KATRIN (Karlsruhe Tritium Neutrino) experiment investigates the energetic endpoint of the tritium beta-decay spectrum to determine the effective mass of the electron anti-neutrino. The collaboration has reported a first mass measurement result at this TAUP-2019 conference. The TRISTAN project aims at detecting a keV-sterile neutrino signature by measuring the entire tritium beta-decay spectrum with an upgraded KATRIN system. One of the greatest challenges is to handle the high signal rates generated by the strong activity of the KATRIN tritium source while maintaining a good energy resolution. Therefore, a novel multi-pixel silicon drift detector and read-out system are being designed to handle rates of about 100~Mcps with an energy resolution better than 300~eV (FWHM). This report presents succinctly the KATRIN experiment, the TRISTAN project, then the results of the first 7-pixels prototype measurement campaign and finally describes the construction of the first TRISTAN module composed of 166 SDD-pixels as well as its implementation in KATRIN experiment.
\end{abstract}

\section*{Introduction}
Great discoveries on the nature of the neutrino have been brought to light by solving experimental anomalies. If we know now that out of three mass eigenstates, two must be non zero, the mechanism generating the neutrino mass is still pending. Most hypotheses need to invoke the right-handed partner of the already observed left-handed active neutrino. This approach could provide a viable Dark Matter candidate with a minimal extension of the standard model via creation of keV-sterile neutrino~\cite{ASAKA}. The mass-range of this sterile neutrino sweeps along several orders of magnitude. Nuclear physics, astrophysics as well as cosmology all provide constraints on the mixing angle of such a particle. The TRISTAN project focuses on the 1-20 keV sterile neutrino candidates where the boundary imposed by astrophysics on the sterile-active mixing angle $\theta$ is on the order of $\sin^2\theta < 10^{-6}$~\cite{2019Boyarsky}. 

The \textsc{KA}rlsruhe \textsc{TRI}tium \textsc{N}eutrino (KATRIN) experiment aims for a direct neutrino mass determination using gaseous tritium. The $\beta$ spectrum of tritium is measured 40-eV around the Q$_\beta$ where the effect of the neutrino mass on the $\beta$ model is maximum. The final sensitivity of KATRIN to the neutrino mass is about 200~meV at 90\%~C.L. after 3~years of data taking~\cite{DesignRep}. To reach such a low sensitivity, the electrons are adiabatically guided in ultra-high vacuum through magnetic field lines from the windowless gaseous tritium source toward a Magnetic Adiabatic Collimator Electrostatic (MAC-E) filter. 
In this configuration, the detector at the end of the line counts electrons with sufficient energy to overcome the electrostatic high voltage (HV). The HV is measured to the ppm precision level using a custom-made HV divider~\cite{Thummler2009} and a commercial digital voltmeter. 
In addition, to control small voltage drifts on long term measurements the former Mainz neutrino mass experiment spectrometer has been coupled to the main spectrometer. It monitors the main spectrometer HV by scanning $^{83m}$Kr conversion electron lines~\cite{Erhard} and is called KATRIN Monitor Spectrometer. It will be used for the first TRISTAN module characterization.

KATRIN reported the first result and the best limit, as of today, on the neutrino mass by direct measurement in this conference~\cite{KATRINmass}. After completion of the neutrino mass program, KATRIN will search for keV-sterile neutrino signature in the full energy range of the tritium spectrum. To reach the required precision, one needs to accumulate statistics as well as master any systematics to the ppm-level. Combining an integral and differential measurement mode is a promising way to mitigate detector/transport effects\cite{Mertens_wavelent, BrunstTroitsk}. All these considerations lead to the development of a new detector and read-out system, specifically designed to hunt a keV sterile neutrino with KATRIN, the TRISTAN project.

\section{The TRISTAN project}
The TRISTAN detector is developed to measure the full tritium $\beta$-decay spectrum at the end of the KATRIN beam line with a resolution better than 300~eV at 18~keV. In this mode of operation, a rate of about $10^{10}$ electrons per second in a 20~cm beam tube must be handled. The technology suited for this purpose is the Silicon Drift Detector (SDD). 
\subsection*{Prototypes}

The first generation of SDD detector has been using monolithic CMOS charge preamplifier connected through wire bonds to the anode of the SDD pixel to reduce capacitance therefore noise. The number of drift rings and the geometry of the pixels have been optimized. Detectors have been tested using x-ray sources and the performance reach our requirements: a resolution of 140~eV (FWHM at 5.9 keV) with a peaking time of 1~$\mu$s~\cite{2019_Mertens}. 

TRISTAN aiming at precisely measuring $\beta$ radiation, the detector response to electrons must be clearly understood. One of the main differences with respect to x-rays is the impact of the entrance window dead-layer. This dead-layer is due to protective silicon oxide layer as well as implantation of acceptors on the silicon bulk. It is a zone with partial collection efficiency, leading to a shift and a distortion of the electron response. Two different techniques were developed using a mono-energetic electron beam to estimate the size of this dead-layer. One is artificially increasing the dead-layer seen by the electron by tilting the detector with respect to the beam axis, the second technique uses Bremsstrahlung radiation on an ultra-thin aluminum foil for an absolute energy shift measurement. This allowed to estimate the dead-layers of our prototype detectors and to properly characterize the electron response. Besides, new implantation techniques are being tested to further reduce this dead-layer showing promising results.

Finally, over the last two-years, 7-pixel SDD prototypes were successfully operated at the Troitsk $\nu$-mass experiment as main detector deriving the first constraints on keV-sterile neutrino using the TRISTAN technology~\cite{BrunstTroitsk, Konrad}. Besides, a TRISTAN prototype has been successfully implemented this year in KATRIN as rate monitoring device.

\subsection*{Final detector}
\begin{figure}[h]
\begin{minipage}{0.5\textwidth}
\centering\includegraphics[width=1.1\textwidth]{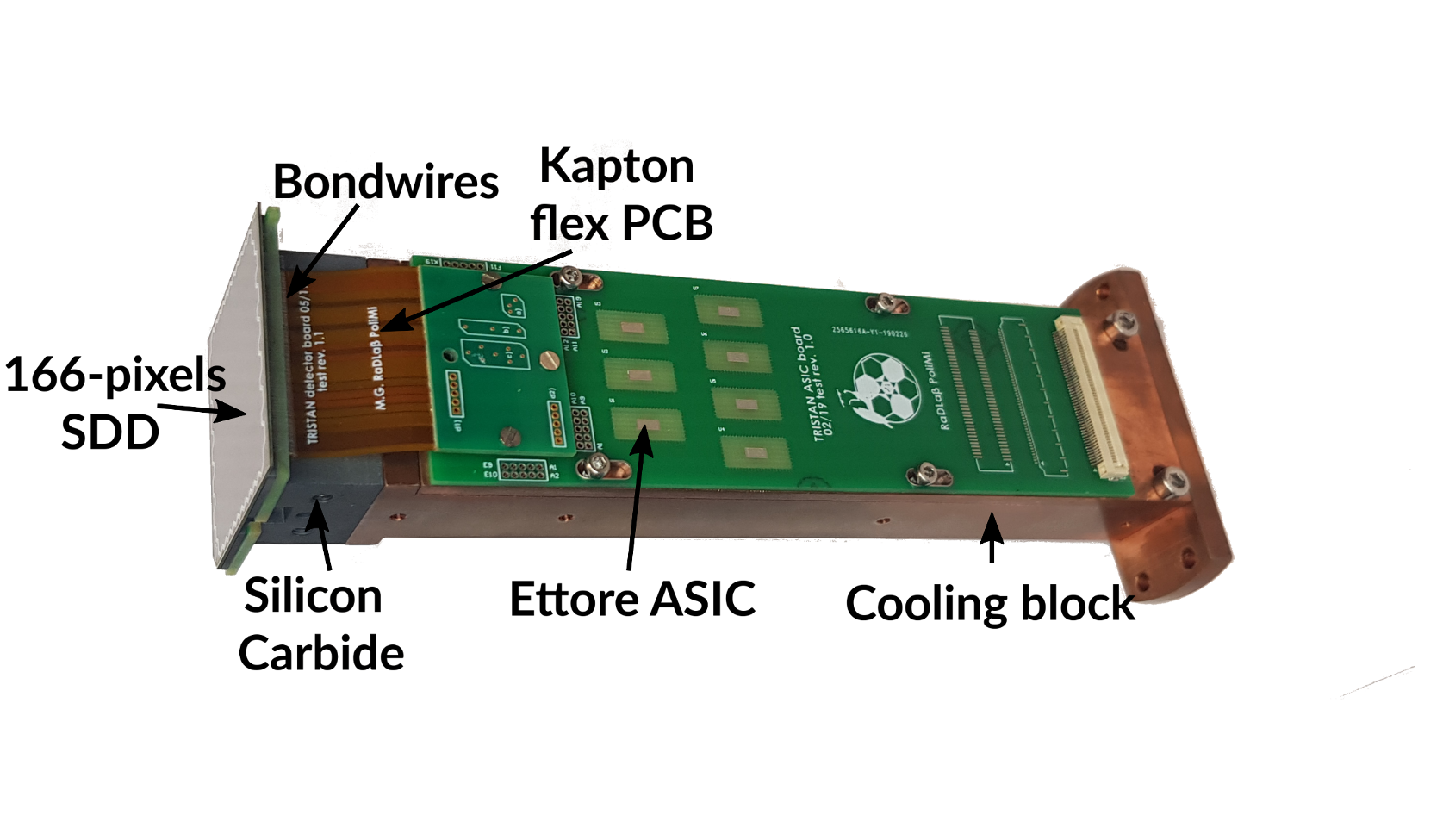}
\caption{Picture of the TRISTAN detector module dummy ($4\times 4 \times 16$ cm$^3$) produced to investigate thermal coupling and mounting assembly procedure.}
\label{fig_dummy}
\end{minipage}
\begin{minipage}{0.05\textwidth}
\hfill
\end{minipage}
\begin{minipage}{0.45\textwidth}
\centering\includegraphics[width=0.65\textwidth]{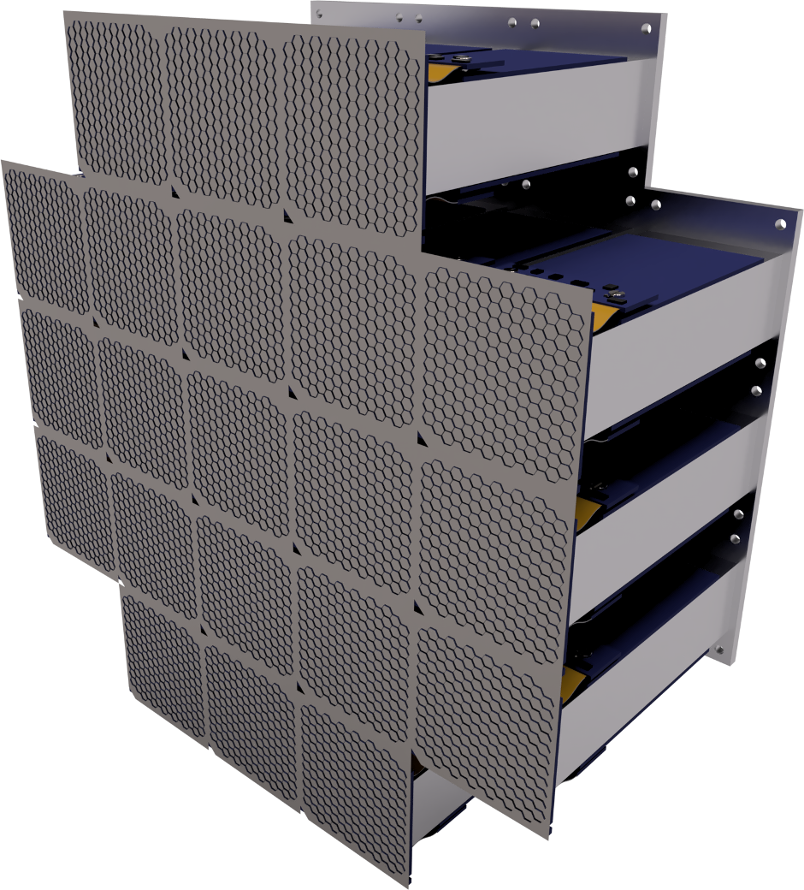}
\caption{Drawing of the TRISTAN detector design: 3486 pixels on a 400~cm$^2$ plane divided into 21 identical modules. }
\label{fig_final}
\end{minipage}
\end{figure}

The KATRIN detector chamber and read-out system will be upgraded to undergo the keV-sterile program with a TRISTAN system after completion of the mass measurement campaign. The total number of pixels as well as the size of the final detector must follow the requirements:
\begin{itemize}\itemsep0pt
\item minimize pile-up events by increasing the number of pixels,
\item avoid charge sharing between adjacent pixel by increasing the size of pixels,
\item reduce backscattering by maximizing transverse momentum,
\item reduce back reflection into different pixels by reducing Larmor radius,
\item keep a manageable complexity.
\end{itemize}
All these constraints have been simulated and an optimal configuration was found with 3486~hexagonal pixels of 3~mm diameter. The detector plane will be split into 21 module disposed as represented in figure~\ref{fig_final}. To reach this target, the TRISTAN group has decided to first focus on the production of one of these modules and to implement it as KATRIN Monitor Spectrometer's detector. The demonstration of the module performance in real conditions will serve as a milestone for the project. It will also improve the stability of the Monitor Spectrometer and allow for new features like picturing the activity distribution of $^{83m}$Kr in real time. 

\section{The first TRISTAN module}
\begin{figure}[h]
\centering\includegraphics[width=\textwidth]{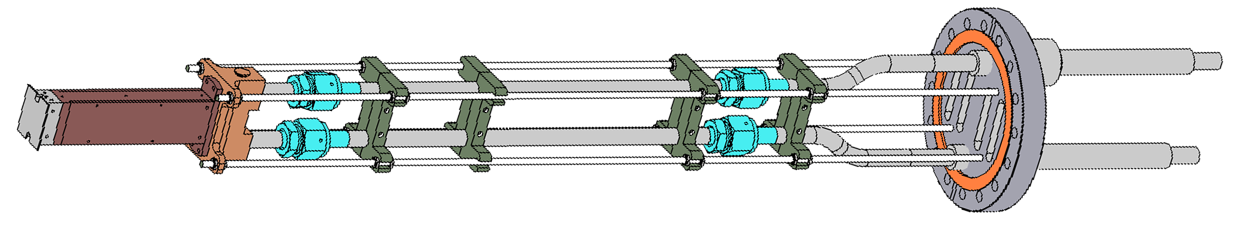}
\caption{First TRISTAN module mechanical set-up. From left to right: SDD, carbon-fiber reinforced silicon carbide (grey), Copper block (red), heat exchange (orange), cooling pipes (light grey), CF100 flange (dark grey).}
\label{fig_mospipe}
\end{figure}

The first module of TRISTAN is a 16~cm high, $4\times 4$~cm$^2$ large column. On top of it, is the $3.8 \times 4.0$~cm$^2$ silicon chip made of 166 3-mm diameter pixels. This is produced by the Max Planck Semiconductor laboratory and will be delivered in November 2019. The SDD chips are designed with integrated junction field effect transistor enabling for a first amplification stage as close as possible from the anode. The 450~$\mu$m-thick silicon chip as well as a thin detector board are then glued to a custom designed carbon-fiber reinforced silicon carbide. This piece, screwed to a copper block, enables to cool down the detector to -50$^\circ$C to reduce leakage current, while avoiding thermal expansion stress. The SDD and electronics are cooled down using circulation of cooling liquid inside vacuum through cooling pipes. Finally, a heat exchange block connect the cooling circuit with the module as represented in figure~\ref{fig_mospipe}.

The connection between SDD back-plane and the detector board is made through two-line wire bonds. The detector board is then connected through a custom-designed multi-layer dense kapton flex cable to the second amplification stage, the ASIC board, located on the side of the copper block. The 12-channel ASIC used in the second amplification stage, Ettore, is developed specifically for TRISTAN. Considering spatial constraints of the final design, the electronics must lie behind the silicon plate to allow several module to aggregate in final design as displayed in figure~\ref{fig_final}. Figure~\ref{fig_dummy} shows the first dummy module built in Munich to test thermal connections as well as module assembly procedure.

The module is then inserted in the Monitor Spectrometer detector chamber. The detector itself is sitting 1-m away of the vacuum-air connection, inside a 6-T super-conductive magnet. This lead to nonmagnetic materials as well as mechanical structure to extract heat and hold the detector inside a 1~m long bellow tube. Figure~\ref{fig_mospipe} shows the actual design of this first TRISTAN module as planed to be used in the KATRIN Monitor Spectrometer. 
\vspace{-0.5cm}

\section*{Conclusion}
Sterile neutrino hypotheses can be investigated using the KATRIN experiment. KATRIN is now taking data for measuring the neutrino mass and first results have recently been published~\cite{KATRINmass}. Looking for keV sterile neutrino would imply to develop a new detector and read-out system which is the TRISTAN project. The KATRIN/TRISTAN project has the objective to reach limits derived by astrophysical observations. To do so, new detectors have been characterized~\cite{2019_Mertens} and successfully implemented in existing MAC-E filter as the Troitsk spectrometer, already settling limits to keV sterile-to-active neutrino mixing angle~\cite{BrunstTroitsk}. The first TRISTAN module of 166 pixels, 1/21 of the final TRISTAN detector is now under production and will be used as KATRIN Monitor Spectrometer detector. This set-up will enable extensive tests of the TRISTAN detector concerning electron response, electronics linearity, outgasing rate, etc while improving performance of the KATRIN monitoring device. 
\section*{References}
\bibliography{bib}


\end{document}